\documentclass[12pt]{article}
\usepackage{color}
\usepackage{amsfonts}
\usepackage{latexsym}
\usepackage{amsmath,amssymb}
\usepackage{verbatim}
\usepackage{
cite}

\newcommand{\bea}{\begin{eqnarray}}
\newcommand{\eea}{\end{eqnarray}}
\newcommand{\be}{\begin{equation}}
\newcommand{\ee}{\end{equation}}
\newcommand{\ba}{\begin{array}}
\newcommand{\ea}{\end{array}}

\def\nn{\nonumber}
\def\half{{1\over2}}
\def\p{\partial}
\def\eps{\epsilon}

\numberwithin{equation}{section}

\def\e {\epsilon}
\def\s{\sigma}
\definecolor{Wei}{rgb}{0.65,0.0,0}
\definecolor{Geo}{rgb}{0.1,0,0.75}

\begin{document}
\vspace{8cm}
\begin{center}
\vspace{8cm}
{ \LARGE {{Chiral Liouville Gravity }}}

\vspace{1cm}

Geoffrey Comp\`ere$^{*\diamond}$, Wei Song$^*$ and Andrew Strominger$^{*\dagger}$

\vspace{0.8cm}

{\it  $*$Center for the Fundamental Laws of Nature, Harvard University,\\
Cambridge, MA, USA}

{\it  $\diamond$ Physique Th\'eorique et Math\'ematique, Universit\'e Libre de Bruxelles,\\
 Bruxelles, Belgium}

{\it  $^\dagger$Radcliffe Institute for Advanced Study, 
Cambridge, MA, USA}

\vspace{0.5cm}
\today 

\vspace{0.5cm}

\vspace{1.0cm}

\end{center}

\begin{abstract}
 Classical two-dimensional Liouville gravity is often considered in conformal gauge which has a residual left and right Virasoro symmetry algebra.  We consider an alternate, chiral, gauge which has a residual right Virasoro Kac-Moody algebra, and no left Virasoro algebra. The Kac-Moody zero mode is the left-moving energy. 
  Dirac brackets of the constrained Hamiltonian theory are derived,  and the residual symmetries are shown to be generated by integrals of the conserved chiral currents. The central charge and Kac-Moody level are computed. The possible existence  of a corresponding quantum theory is discussed.

\end{abstract}
\thispagestyle{empty}

\pagebreak
\setcounter{tocdepth}{3}

\tableofcontents

\thispagestyle{empty}
\def\cR{{\cal R}}
\section{Introduction}
 Critical behavior  in two spacetime dimensions is often associated with the emergence of an infinite-dimensional
 conformal symmetry group \cite{Belavin:1984vu}. These  symmetries act infinitesimally on the spacetime coordinates $t^\pm$ as
 \be \label{ne}\delta t^+=\e^+(t^+), ~~~~~~~\delta t^-=\e^-(t^-).\ee
 The Lie bracket algebra of these symmetries is two copies of the (centerless) Virasoro algebra.
 Such critical behavior has powerful  consequences and arises in a wide variety of physical and mathematical systems.
 In the last few years, indications have been accumulating \cite{Anninos:2008fx,Compere:2008cv,Guica:2008mu,Castro:2009jf,Compere:2009zj, Compere:2009qm,Guica:2010sw,Hofman:2011zj,ElShowk:2011cm,Song:2011sr,Guica:2011ia,Azeyanagi:2012zd,Detournay:2012pc} of another possible type of two-dimensional critical behavior with an infinite-dimensional symmetry group, sometimes referred to as warped conformal symmetry. Warped conformal symmetries act infinitesimally on the spacetime coordinates $t^\pm$ in the chiral fashion
 \be\label{wo} \delta t^+=\e(t^+), ~~~~~~~\delta t^-=\s(t^+).\ee
 The Lie bracket algebra here consists of one  chiral Virasoro-Kac-Moody algebra. The Virasoro zero mode generates right translations while the Kac-Moody zero mode generates left translations.  So far there has been no clear example in which this critical behavior is nontrivially realized. In this paper we address this issue by the construction of a chiral analog of Liouville theory with warped conformal symmetry at the semiclassical level. The full quantum problem is left for future work.

 The standard conformal symmetry (\ref{ne}) emerges in both gravitational and non-gravitational systems.\footnote{In non gravitational systems, it emerges as an enhanced symmetry at a critical point, while in gravitational systems it is a subgroup of the diffeomorhpisms which acts nontrivially.} We consider here the gravitational context. Theories of gravity by definition transform covariantly under all diffeomorphisms.  In conformal gauge the metric is required to obey
 \be \label{za} g_{--}=g_{++}=0. \ee
 This condition is preserved by the subgroup of two-dimensional diffeomorphisms of the form (\ref{ne}).
This implies \cite{David:1988hj,Distler:1988jt} that the resulting gravity theory is equivalent to a conformal field theory without gravity  in which the unconstrained component of the metric $g_{+-}$ serves as  one of the fields.

Instead of (\ref{za}) we here impose ``chiral gauge"\footnote{This gauge is similar to, though different than, the light cone gauge $g_{--}=0=\p_\pm g_{+-}$ adopted in KPZ \cite{Knizhnik:1988ak}, who found  a chiral $SL(2,R)$  Virasoro-Kac-Moody. Our Kac-Moody current ($j_\s$ below) is closely related to one of the KPZ $SL(2,R)$ currents. The relation of our results to theirs remains to be explored.}  \be \label{wnd}g_{--}=0,~~~~~ \p_-\big(g^{+-}g_{++}\big)=0.
\ee
This condition is violated by the usual conformal symmetries (\ref{ne}), but preserved by the warped conformal symmetries (\ref{wo}). This implies  that the resulting gravity theory is equivalent to a warped conformal field theory without gravity  in which the unconstrained components of the metric serve as fields.  We work out the details for the example of Liouville gravity (see \cite{Jackiw:1982hg, Seiberg:1990eb,Teschner:2001rv}), beginning from the covariant Polyakov action \cite{Polyakov:1987zb}. Gauge fixing this covariant theory results in a chiral version of the standard Liouville theory.

This paper is organized as follows. In section 2 we review the nonlocal but covariant Polyakov action for Liouville gravity. In section 3 we impose the chiral gauge condition and derive a local action with constraints. The two Noether currents associated to the residual Kac-Moody-Virasoro are shown to be conserved, chiral and given by appropriate components of the stress tensor.  The theory is shown to be integrable via a B\"acklund transformation to free fields.
Section 4 gives the Hamiltonian formalism on surfaces of constant $t^-$.  Dirac brackets following from  the constraints are constructed. In principle the quantum theory is defined by by replacing these brackets with quantum commutators. The charges are shown, under the Dirac bracket,  to obey a Kac-Moody-Virasoro algebra and the central charge and level are computed. In section 5 an infinite class of generalizations involving matter are described.  Section 6 closes with comments on the quantum theory.

In a holographically dual companion paper \cite{css1}, we consider the problem of  $SL(2,R)_L\times    SL(2,R)_R$ Chern-Simons gravity on AdS$_3$. It is well known \cite{Coussaert:1995zp} that, with the usual Brown-Henneuax AdS$_3$ boundary conditions, this is dual to ordinary Liouville theory on the boundary. We find a new, chiral set of AdS$_3$ boundary conditions for which
the asymptotic symmetry group is generated by a Virasoro-Kac-Moody algebra.  Applying these boundary conditions to Chern-Simons gravity on AdS$_3$ gives exactly the chiral Liouville theory discussed in this paper. Hence the bulk and boundary  analyses are compatible.

\section{Covariant action}
Two dimensional  Liouville gravity can be described by the nonlocal Polyakov action \cite{Polyakov:1987zb}
\be S^0_L={c \over 96 \pi}\int d^2x(Z \cR -2 \Lambda\sqrt{-g}),\ee
where
\bea \cR &\equiv & \sqrt{-g} R, \cr
Z(x) &\equiv& \int d^2x'G(x,x')\cR (x'),\cr
~~~~\sqrt{-g}g^{ab}\nabla_a\nabla_bG(x,x')&=&\delta^2(x,x') .\eea
The stress tensor defined by the metric variations of this action is\footnote{ We have used the relations $
\delta \cR = \sqrt{-g}\nabla^a (\nabla^b \delta g_{ab} - g^{bc}\nabla_a \delta g_{bc})$ and $
\delta G(x,x')=- \int d^2 x'' G(x,x') \delta (\sqrt{-g}\nabla^2)G(x,x'').
$}
\bea
 {2\pi\over\sqrt{-g}}{\delta S_L^0 \over \delta g^{ab}}\equiv&& T^0_{ab}\nn\\ = &&\frac{c}{48}\big(2g_{ab}\nabla^2Z-2\nabla_a \nabla_bZ +  \Lambda g_{ab}\nn\\
&&~~~~+\nabla_a Z \nabla_bZ - \half g_{ab}\nabla_cZ \nabla^cZ\bigr).\label{Tp}
\eea
The trace of the stress tensor  is the local expression
\be T^{0a}_{~~a}= \frac{c}{24}(R+\Lambda).\ee

\section{Chiral gauge}

\subsection{Gauge condition}
Liouville gravity is often considered in conformal gauge where $Z$ reduces to the logarithm of the conformal factor,   $S^0_L$ is local
and the residual symmetry algebra consists of a left and a right Virasoro.  Here we consider instead chiral gauge
\be \label{wn}g_{--}=0,\ee \be \p_-\big(g^{+-}g_{++}\big)=0.
\label{chiralg} \ee
It is easy to see any metric can be locally put into chiral gauge. (\ref{wn}) implies the metric can be written as
\be \label{met} ds^2=-e^{2 \rho}(dt^+dt^--h (dt^+)^2).\ee
Imposing  (\ref{chiralg}) further implies
\be  \p_-h=0.\ee
\subsection{Residual symmetries}
The full gauge conditions (\ref{wn}) and (\ref{chiralg}) leave unfixed  a residual symmetry generated by two right-moving ($t^+$-dependent) functions
\be \label{ct} \delta t^+=\e(t^+),~~~\delta t^-=\sigma (t^+).\ee
The symmetry algebra is a right-moving  Virasoro Kac-Moody. The Kac-Moody zero mode is left translations.
Under the residual symmetries, the fields transform as
\bea\label{res}
\delta \rho &=& -\frac{1}{2}\p_+ \e -\e\p_+ \rho -\s\p_- \rho,\\
\delta h  &=&\p_+ \s - h  \p_+ \e -\e\p_+ h -[\s\p_-h]\, .
\eea
Here and below we write in  square brackets terms which vanish only when the second gauge condition $\p_-h=0$ in (\ref{chiralg}) is imposed.

(\ref{chiralg}) eliminates the usual left-moving Virasoro. However the left translational zero mode remains. It is convenient to study the theory in a fixed sector for this zero mode. This is facilitated  by adding to the action a term
\be \label{dl} S_L=S_L^0+\frac{\Delta}{4\pi}\int d^2x \sqrt{-g} g^{--}= S_L^0-{\Delta\over2\pi} \int d^2x h.\ee
The $g^{--}$ equation of motion then gives
\be T_{--}=T^0_{--}+\frac{\Delta}{2}=0 \label{T--} \ee
where the constant $\Delta/2$ is the left-moving energy density.

\subsection{Geometry}
 The nonzero connection coefficients for a metric of the form (\ref{met}) are  \bea
\Gamma^+_{++}&=&2 \p_+\rho+2 h \p_-\rho+[\p_- h],\cr
\Gamma^-_{--}&=&2 \p_-\rho,\cr
\Gamma^-_{-+}&=&-2 h \p_-\rho-[\p_- h],\cr
\Gamma^-_{++}&=&2 h \p_+\rho-\p_+h +4 h ^2\p_-\rho+[2h\p_- h],
\eea
and the Ricci density is
\be \cR=4\p_- \p_+\rho+4h \p_-^2\rho +[4\p_- h \p_-\rho +{2} \p_-^2 h] = -2\sqrt{-g} \nabla^2 \rho+[2\p_-^2 h]. \ee
 $Z$ is then
\be Z(x)=-2 \rho(x)+[2\int d^2x' G(x,x')\p_-^2 h(x')].\ee
The action reduces to (up to boundary terms)
\bea
S_L &=& \frac{c}{12\pi }\int d^2x \Big(\p_+\rho \p_-\rho -\frac{\Lambda}{8} e^{2 \rho} +  h  (\p_-\rho)^2-{6\over c} h\Delta \nn\\
&&+[\p_-h\p_- \rho+\half \int d^2x' \p_-^2 h(x) G(x,x')\p_-^2 h(x')]\Big).\label{action2}
\eea

\subsection{Stress tensor and Noether current}

The stress-tensor \eqref{Tp} is, when $\p_-h=0$,
\bea T^0_{++} &=& -\frac{c}{12}\Big((\p_+\rho)^2- \p_+^2 \rho -4h (\p_-\p_+ \rho -\frac{1}{2} \p_+ \rho \p_-\rho)-\frac{\Lambda}{4} h e^{2 \rho}\nn\\
&& -\p_+ h \p_- \rho-4h^2 (\p_-^2 \rho - \frac{1}{2} (\p_-\rho)^2 )\Big),\nn\\
T^0_{+-}&=&-\frac{c }{12}\Big( \p_+\p_- \rho +\frac{\Lambda}{8}e^{2 \rho}+2h \p_-^2 \rho -h (\p_-\rho)^2\Big),\nn \\
T^0_{--} &=& - \frac{c }{12}\Big((\p_-\rho)^2 - \p_-^2\rho\Big).
\eea
The Noether procedure leads to the following conserved currents
\bea
j^-_\epsilon&=&{c\over6}\left((\p_+\rho)^2-\p_+^2\rho+2h(\p_+\rho\p_-\rho-\p_+\p_-\rho)-\p_+h\p_-\rho\right),\\
j^-_\sigma &=& {c\over6}h\Big((\p_-\rho)^2-\p_-^2\rho\Big)+h\Delta,\\
j^+_\sigma &=& \frac{c}{6}\left( (\p_-\rho)^2 -\p_-^2 \rho\right)-\Delta,\\
j^+_\eps &=& - h j^+_\sigma,
\eea
which can be summarized in terms of the total stress tensor,
\bea
T_{ab}=T^0_{ab}+\Delta \left( \begin{array}{cc} h^2 & -\frac{3}{2}h \\ -\frac{h}{2} & \frac{1}{2}\end{array} \right)_{ab}
\eea
as
\bea
j^a_\eps = e^{2\rho}T^a_{\;\;-},\qquad j^a_\sigma = e^{2\rho}T^a_{\;\; +}. \qquad
\eea
Here indices are raised with the full metric (\ref{met}). The full $T_{ab}$ is not symmetric because of the asymmetric function of the metric added to the action in (\ref{dl}).  After imposing the equation $T_{--}=0$, $j^+_\eps = j^+_\sigma =0$ we are left with two right-moving conserved currents
\bea
\p_ - j^-_\eps &=& 0,\nn \\
\p_- j^-_\sigma &=& 0,\label{cons}
\eea
associated to the residual symmetries.

In the familiar conformal gauge treatment, the vanishing of the trace $T_{+-}=0$ is imposed as an equation of motion,  but one cannot set
$T_{--}=0$ or $T_{++}=0$ because this condition is not invariant under the residual Virasoro symmetries.
This can be variously understood as due to the conformal  anomaly or to the fact that the Virasoro subgroup of the diffeomorphisms generates the nontrivial asymptotic symmetry group.
In either case, the weaker conditions $\p_+T_{--}=0$ and $\p_-T_{++}=0$ follow from the trace equation. In our case the conditions
$j^-_\epsilon=0$  or $j^-_\sigma=0$ are not invariant under the residual Virasoro-Kac-Moody and cannot be imposed.
However the trace equation  $R+\Lambda=0$ as well as $T_{--}=0$ are invariant under the residual symmetry and can be imposed. These then imply the chiral conservation laws \eqref{cons}.

\subsection{B\"acklund transformation}
  Chiral Liouville gravity, like its conformal gauge cousin, is integrable. Here we demonstrate this by mapping to a set of free fields. The B\"acklund transformation is
 \bea \label{bk}e^{2\rho}&=&-{8\over\Lambda} \sqrt{6\Delta\over c}e^{2O} \Big(\cosh\big(\sqrt{6\Delta\over c}(t^--P)\big)-\sinh\big(\sqrt{6\Delta\over c}(t^--P)\big)\int e^{2O}\Big)^{-2}, \cr  h&=&\p_+P,\eea
where $\p_+\int e^{2O}=e^{2O}$.  The equations of motion $R=-\Lambda$ and $T_{--}=0$ then reduce to
 the free field equations
 \be \p_-O=0,~~~\p_-P =0. \ee
Evaluating the currents with the ansatze (\ref{bk}) reveals
\bea j^-_\e&=&{c\over 6}\Big((\p_+O)^2-\p^2_+O\Big)-\Delta (\p_+P)^2,\\  j^-_\s&=&2\Delta\delta^{-+} \p_+P. \eea

\section{Canonical formalism}
In this section we use $t^-$ as time and derive  the constrained  Hamiltonian theory. For this purpose we may use the action
\be \label{xq}S_L=\frac{c}{12\pi }\int d^2x\Big(\p_+\rho \p_-\rho -\frac{\Lambda}{8} e^{2 \rho} +  h  (\p_-\rho)^2 +[\p_-h\p_- \rho]-{6\over c}h\Delta \Big).\ee
The term in square brackets cannot be dropped even though it vanishes for $\p_-h$=0 because it affects the linear $h$ variation. However we can drop the last, nonlocal term appearing in (\ref{action2}) because it vanishes quadratically on the constraint submanifold $\p_-h=0$ and hence has no effect on the final formulae.
In the next subsection we give the Poisson brackets for the unconstrained theory with $h=h(t^+,t^-)$. In the following one  we find and impose the constraints and construct the corresponding Dirac brackets. In the last one we verify that the canonical charges generate the residual symmetries.

\subsection{Unconstrained Poisson brackets}
The (rescaled) canonical mometa are defined as
\bea
{12 \pi\over c}{\delta S_L\over \delta \p_-\rho}&\equiv& \Pi_\rho=(\p_+\rho+2h\p_-\rho+\p_-h),\cr \quad {12 \pi\over c}{\delta S_L\over \delta \p_-h}&\equiv& \Pi_h=\p_-\rho.
\eea
This can be inverted to give the time derivatives of the coordinates in terms of the momenta as  \bea \p_-\rho=\Pi_h,\quad \p_-h=\Pi_\rho-\p_+\rho-2h\Pi_h.\eea
The Poisson brackets are \bea \{\rho(t^+),\Pi_\rho(s^+)\}&=&{12\pi\over c}\delta (t^+-s^+),\\  \quad  \{h(t^+),\Pi_h(s^+)\}&=&{ 12\pi\over c}\delta (t^+-s^+),\\ \{h,\Pi_\rho\}=\{h,\rho\}&=&\{\rho,\Pi_h\}=\{\Pi_h,\Pi_\rho\}=0.\eea
The Hamiltonian is
\bea
H = \int dt^+ (\Pi_h \Pi_\rho - \Pi_h \p_+\rho- h \Pi_h^2 +\frac{\Lambda}{8}e^{2\rho}+\frac{\Delta}{k}h).
\eea

\subsection{Constrained Dirac brackets}
Now we restrict to a constrained  submanifold of phase space on which
\bea C_1&\equiv& {\p_-h}=\Pi_\rho-\p_+\rho-2h\Pi_h=0.\eea
The condition $C_1=0$ is not preserved by the equations of motion following from (\ref{xq}), or equivalently does not commute with the unconstrained Hamiltonian. The vanishing of the commutator requires the second constraint
\bea C_2&\equiv& \p_+\Pi_h-{6\over c}\Delta h+h\Pi^2_h+{ \Lambda\over8}e^{2\rho}=0.
\eea
The submanifold of phase space defined by $C_1=C_2=0$ is preserved by unconstrained Hamiltonian evolution.

The Poisson brackets of constraints with themselves, fields and momenta are
\bea
{c \over12\pi}\{C_1(t^+),C_1(s^+)\}&=&-2\p_{t^+}\delta(t^+-s^+),\nn\\
{c \over12\pi}\{C_2(t^+),C_2(s^+)\}&=& \p_{t^+}\delta(t^+-s^+)\Big({12\over c}\Delta-\Pi^2_h(s^+)-\Pi^2_h(t^+)\Big),\nn\\
{c \over12\pi}\{C_1(t^+),C_2(s^+)\}&=&2\p_{t^+}(\Pi_h(t^+)\delta(t^+-s^+))-{24\over c}h\Delta\delta(t^+-s^+),\nn\\
&&-2 C_2(t^+) \delta(t^+-s^+),\nn\\
{c \over12\pi}\{h(t^+),C_1(s^+)\}&=&-2h(s^+)\delta(t^+-s^-),\nn
\\
{c \over12\pi} \{h(t^+),C_2(s^+)\}&=&\p_{s^+}\delta(t^+-s^+)+2h(t^+)\Pi_h(t^+)\delta(t^+-s^+),\nn\\
{c \over12\pi} \{\rho(t^+),C_1(s^+)\}&=& \delta(t^+-s^+),\nn\\
 \{\rho(t^+),C_2(s^+)\}&=&0,\nn\\
{c \over12\pi}\{\Pi_h(t^+),C_1(s^+)\}&=&2\Pi_h(s^+)\delta(t^+-s^-),\nn
\\
{c \over12\pi} \{\Pi_h(t^+),C_2(s^+)\}&=&({6\over c}\Delta-\Pi_h^2(t^+))\delta(t^+-s^+),\nn\\
{c \over12\pi} \{\Pi_\rho(t^+),C_1(s^+)\}&=&- \p_{t^+}\delta(t^+-s^+),\nn\\
{c \over12\pi} \{\Pi_\rho(t^+),C_2(s^+)\}&=&-{\Lambda\over4}e^{2\rho}\delta(t^+-s^+).
\eea
The general formula for Dirac brackets is \be [A,B]=\{A,B\}-\{A,C_i\}\{C_i,C_j\}^{-1}\{C_j,B\}.\ee
In our case the quantities   \be b^{ij}(t,s)\equiv {12 \pi \over c}\{C_i(s),C_j(t)\}^{-1}
\ee
are determined by the partial differential equations
\bea
\delta(t-w)&=&
-2\p_t\Big(b^{11}(t,w)-\Pi_h(t) b^{21}(t,w)\Big)-{24\over c}\Delta h(t)b^{21}(t,w),
\nn\\
\delta(t-w)&=&{12\Delta\over c\Pi_h(t)}\Big(\p_t b^{21}(t,w)+2 h(t)\big({b^{11}(t,w)}-\Pi_h(t)b^{21}(t,w)\big)\Big),\nn\\
\delta(t-w)&=&{12\Delta\over c}\Big(\p_t b^{22}(t,w)+2h(t)\big(b^{12}(t,w)-\Pi_h(t)b^{22}(t,w)\big)\Big),\nn\\
0&=&\p_t\Big(b^{12}(t,w)-\Pi_h(t) b^{22}(t,w)\Big)+{12\over c}\Delta h(t)b^{22}(t,w).
\label{pde}\eea
The relations  \bea \int ds \{h(t), C_i(s)\}b^{i1}(s,w)&=&-{\pi \delta(t-w)\over \Delta}\Pi_h(t) ,\cr \int ds \{h(t), C_i(s)\}b^{i2}(s,w)&=&-{\pi \delta(t-w)\over \Delta} \eea
imply the simplification of Dirac brackets involving $h$
\bea  [h(t^+),h(s^+)]
&=&-{\pi \over \Delta }\p_{t^+}\delta(t^+-s^+),\\
\,[h(t^+),\rho(s^+)]&=&-{\pi\over\Delta}\delta(t^+-s^+)\Pi_h(s^+)=-{\pi\over\Delta}\delta(t^+-s^+)\p_-\rho(s^+),\nonumber\\
\,[h(t^+),\Pi_h(s^+)]&=&-{\pi\over\Delta}(\Pi_h^2-{6\over c}\Delta)\delta(t^+-s^+)=-{\pi\over\Delta}\p_-^2\rho\delta(t^+-s^+),\nonumber\\
\,[h(t^+),\Pi_\rho(s^+)]&=&{\pi \Lambda\over4\Delta}e^{2\rho}\delta(t^+-s^+)-{\pi\over\Delta}\Pi_h(t^+)\p_{t^+}\delta(t^+-s^+).\nonumber\\
\eea
The Dirac bracket containing $\rho$, however, requires solving the partial differential equation system  (\ref{pde}), \be \,[\rho(t^+),\rho(s^+)]=b^{11}(t^+,s^+).\ee
In weak field perturbation theory this reduces at leading order to
\be\,[\rho(t^+),\rho(s^+)]=-\half\Theta(t^+-s^+), \ee
with $\Theta$ the sign function.
\subsection{Charges}
The canonical forms of the  Noether currents are, when the constraints are imposed
\bea
j^-_\epsilon&=&{c\over6}\Big(\Pi_\rho\p_+\rho-h\p_+\Pi_h-{1\over2}\p_+\Pi_\rho-{1\over 2}\p_+^2\rho\Big),\\
j^-_\sigma&=&2\Delta h,\\ j^+_\e&=&j^+_\s=0.\eea
These generate the residual symmetries under the Dirac bracket
\bea{1\over2\pi}  [\int dt^+(\e j^-_\e+\s j^-_\s), \rho]&=&-\frac{1}{2}\p_+ \e -\e\p_+ \rho -\s\p_- \rho, \cr{1\over2\pi} [\int dt^+(\e j^-_\e+\s j^-_\s), h] &=& \p_+ \s - h  \p_+ \e -\e\p_+ h. \eea
Using the properties that
\bea
\int ds^+ \{ j^-_\eps(t^+),C_i(s^+) \}b^{ij}(s^+,w^+) =0, \qquad j=1,2,
\eea
the current-current  commutators  are obtained as
\bea [ j^-_\sigma(t^+),j^-_\sigma(s^+)]&=&{\pi k\, }\p_{t^+}\delta(t^+-s^+),\cr
 \,[ j^-_\epsilon(t^+), j^-_\sigma(s^+)]&=&{2\pi j^-_\sigma(t^+)\p_{t^+}\delta(t^+-s^+)} \,\cr  [ j^-_\epsilon(t^+), j^-_\epsilon(s^+)]&=&2\pi \p_{t^+}\delta(t^+-s^-)(j^-_\epsilon(t^+)+j^-_\epsilon(s^+))\cr&&~~~~~-{\pi c\over6}\p_{t^+}^3\delta(t^+-s^+). \eea
 This is a Virasoro Kac-Moody algebra with central charge $c$ and level
 \be k=-4\Delta.\ee
 We note that the sign of the level is the opposite of the sign of $\Delta$, and in particular flips at $\Delta=0$.
 \section{Coupling to matter}
   As discussed in action 3, the warped conformal symmetry  arises as a residual subgroup of two-dimensional diffeomorphisms in chiral gauge.  Therefore an infinite  class of classical warped conformal field theories can be obtained simply by starting with any generally covariant theory (with a metric) in two dimensions, gauge fixing to chiral gauge and fixing $T^{matter}_{--} = \Delta/2$. This is similar to the procedure \cite{David:1988hj,Distler:1988jt} which generates ordinary conformal field theories from two-dimensional generally covariant theories.

Let us illustrate it with a massive scalar field theory coupled to gravity in two dimensions. The covariant action is
\be S=S^0_L-\half\int d^2x \sqrt{-g}(g^{ab}\p_a\phi\p_b \phi+m^2\phi^2).\ee
In chiral gauge this becomes
\be\label{pij} S=S^0_L+\int dt^+dt^-(\p_+\phi\p_- \phi+h\p_-\phi\p_- \phi - {m^2\over 4}e^{2\rho}\phi^2),\ee
from which one can derive the corrections to the currents and Dirac brackets. This has many obvious generalizations.

We note that (\ref{pij}) differs from the usual conformal gauge result only by the middle term.  It may be possible to think of the chiral theory as a usual conformal field theory deformed by this term. In conformal field theory language $h$, being proportional to a right moving current, is dimension $(1,0)$, while $(\p_-\phi)^2$ is dimension $(0,2)$. Hence the middle term resembles a dimension $(1,2)$ operator. Such deformations indeed have very special properties.  They have been studied in the context of warped conformal field theories and the Kerr/CFT correspondence in \cite{Guica:2010sw,Compere:2010uk,ElShowk:2011cm} and are related to IR limits of the dipole deformed gauge theories  \cite{Bergman:2000cw,Dasgupta:2000ry,Bergman:2001rw}. This connection merits further investigation.

 \section{Quantum theory}
 In this paper we have described the classical Liouville gravity in chiral gauge, found the residual symmetry, shown that it is classically soluble and constructed the currents and Dirac brackets. The status of chiral Liouville  theory hence is roughly that of ordinary conformal gauge Liouville theory three decades ago \cite{Curtright:1982gt,D'Hoker:1982er,Gervais:1983am,D'Hoker:1983is}. In the intervening period, ordinary Liouville theory has been found to  have an extraordinarily rich and beautiful structure at the quantum level, see $e.g.$ \cite{Zamolodchikov:1995aa,Fateev:2000ik, Teschner:2000md,Teschner:2001rv,Alday:2009aq}. It is natural to try to construct a quantum version of chiral Liouville theory. In principle, the quantum theory is defined by turning the Dirac brackets of section 4 into quantum commutators.  However because of the chiral nature of the residual symmetry, it is not obvious that a regulator can be found which preserves the symmetry or that the quantum theory exists at all. If it does exist, it should have a structure comparable in richness to the ordinary Liouville theory. Perhaps there is a transformation which maps the ordinary Liouville correlators to those of the chiral Liouville theory. We leave this issue to future investigations.
\section*{Acknowledgements}
This work was supported by NSF grant 1205550.
W.S. is supported in part by the Harvard Society of Fellows, G.C. is a Research Associate of the Fonds de la Recherche Scientifique F.R.S.-FNRS (Belgium) and A.S is a Fellow at the Radcliffe Institute for Advanced Study.


\providecommand{\href}[2]{#2}\begingroup\raggedright\endgroup

 \end{document}